\documentclass[twoside,9pt]{article}
\usepackage{extsizes}
\usepackage[super,sort&compress,comma]{natbib}
\usepackage[version=3]{mhchem}
\usepackage[left=1.5cm, right=1.5cm, top=1.785cm, bottom=2.0cm]{geometry}
\usepackage{balance}
\usepackage{times,mathptmx}
\usepackage{sectsty}
\usepackage{graphicx}
\usepackage{lastpage}
\usepackage[format=plain,justification=justified,singlelinecheck=false,font={stretch=1.125,small,sf},labelfont=bf,labelsep=space]{caption}
\usepackage{float}
\usepackage{fancyhdr}
\usepackage{fnpos}
\usepackage[english]{babel}
\usepackage{array}
\usepackage{droidsans}
\usepackage{charter}
\usepackage[T1]{fontenc}
\usepackage[usenames,dvipsnames]{xcolor}
\usepackage{setspace}
\usepackage[compact]{titlesec}
\usepackage{soul}
\definecolor{cream}{RGB}{222,217,201}

\usepackage{amsmath}
\usepackage{amssymb}
\usepackage{lipsum}
\usepackage{bm}
\usepackage{graphicx}

\begin{document}

\pagestyle{fancy}
\thispagestyle{plain}
\fancypagestyle{plain}{

%%%HEADER%%%
\fancyhead[C]{\includegraphics[width=18.5cm]{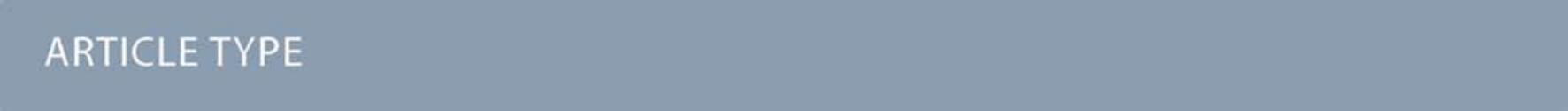}}
\fancyhead[L]{\hspace{0cm}\vspace{1.5cm}\includegraphics[height=30pt]{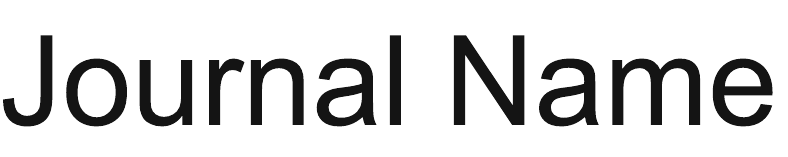}}
\fancyhead[R]{\hspace{0cm}\vspace{1.7cm}\includegraphics[height=55pt]{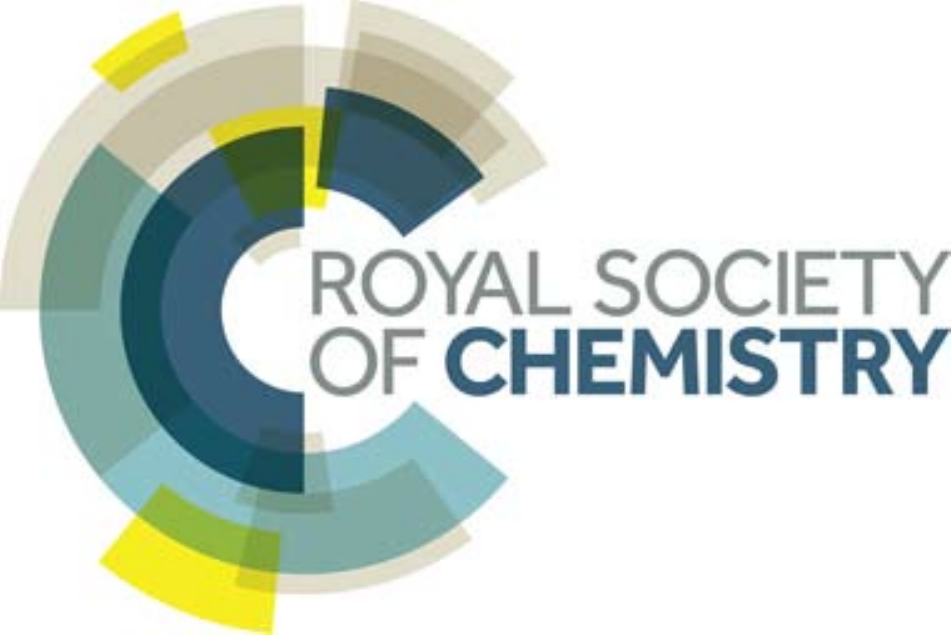}}
\renewcommand{\headrulewidth}{0pt}
}
%%%END OF HEADER%%%

%%%PAGE SETUP - Please do not change any commands within this section%%%
\makeFNbottom
\makeatletter
\renewcommand\LARGE{\@setfontsize\LARGE{15pt}{17}}
\renewcommand\Large{\@setfontsize\Large{12pt}{14}}
\renewcommand\large{\@setfontsize\large{10pt}{12}}
\renewcommand\footnotesize{\@setfontsize\footnotesize{7pt}{10}}
\makeatother

\renewcommand{\thefootnote}{\fnsymbol{footnote}}
\renewcommand\footnoterule{\vspace*{1pt}% 
\color{cream}\hrule width 3.5in height 0.4pt \color{black}\vspace*{5pt}}
\setcounter{secnumdepth}{5}

\makeatletter
\renewcommand\@biblabel[1]{#1}
\renewcommand\@makefntext[1]% 
{\noindent\makebox[0pt][r]{\@thefnmark\,}#1}
\makeatother
\renewcommand{\figurename}{\small{Fig.}~}
\sectionfont{\sffamily\Large}
\subsectionfont{\normalsize}
\subsubsectionfont{\bf}
\setstretch{1.125} %In particular, please do not alter this line.
\setlength{\skip\footins}{0.8cm}
\setlength{\footnotesep}{0.25cm}
\setlength{\jot}{10pt}
\titlespacing*{\section}{0pt}{4pt}{4pt}
\titlespacing*{\subsection}{0pt}{15pt}{1pt}
%%%END OF PAGE SETUP%%%

%%%FOOTER%%%
\fancyfoot{}
\fancyfoot[LO,RE]{\vspace{-7.1pt}\includegraphics[height=9pt]{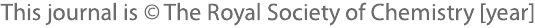}}
\fancyfoot[CO]{\vspace{-7.1pt}\hspace{13.2cm}\includegraphics{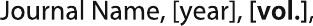}}
\fancyfoot[CE]{\vspace{-7.2pt}\hspace{-14.2cm}\includegraphics{head_foot/RF}}
\fancyfoot[RO]{\footnotesize{\sffamily{1--\pageref{LastPage} ~\textbar  \hspace{2pt}\thepage}}}
\fancyfoot[LE]{\footnotesize{\sffamily{\thepage~\textbar\hspace{3.45cm} 1--\pageref{LastPage}}}}
\fancyhead{}
\renewcommand{\headrulewidth}{0pt}
\renewcommand{\footrulewidth}{0pt}
\setlength{\arrayrulewidth}{1pt}
\setlength{\columnsep}{6.5mm}
\setlength\bibsep{1pt}
%%%END OF FOOTER%%%

%%%FIGURE SETUP - please do not change any commands within this section%%%
\makeatletter
\newlength{\figrulesep}
\setlength{\figrulesep}{0.5\textfloatsep}

\newcommand{\topfigrule}{\vspace*{-1pt}% 
\noindent{\color{cream}\rule[-\figrulesep]{\columnwidth}{1.5pt}} }

\newcommand{\botfigrule}{\vspace*{-2pt}% 
\noindent{\color{cream}\rule[\figrulesep]{\columnwidth}{1.5pt}} }

\newcommand{\dblfigrule}{\vspace*{-1pt}% 
\noindent{\color{cream}\rule[-\figrulesep]{\textwidth}{1.5pt}} }

\makeatother
%%%END OF FIGURE SETUP%%%

%%%TITLE, AUTHORS AND ABSTRACT%%%
\twocolumn[
  \begin{@twocolumnfalse}
\vspace{3cm}
\sffamily
\begin{tabular}{m{4.5cm} p{13.5cm} }

\includegraphics{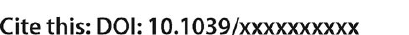} 
& \noindent\LARGE{\textbf{A connection
  between Living Liquid Crystals and electrokinetic phenomena in
  nematic fluids}} \\
\vspace{0.3cm} & \vspace{0.3cm} \\

 & \noindent\large{Christopher Conklin$^{1}$, Jorge Vi\~{n}als$^{2}$, and 
Oriol T. Valls}$^{3}$ \\

\includegraphics{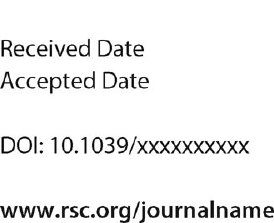} & \noindent\normalsize{We develop a
  formal analogy between configurational stresses in two distinct
  physical systems, and study the flows that they induce when the
  configurations of interest include topological defects. The two
  systems in question are electrokinetic flows in a nematic
  fluid under an applied electrostatic field, and the motion of self
  propelling or active particles in a nematic matrix (a living
  liquid crystal). The mapping allows the extension, within certain
  limits, of existing results on transport in electrokinetic systems to self
  propelled transport. We study motion induced by a pair of point
  defects in a dipole configuration, and steady rotating flows due to
  a swirling vortex nematic director pattern. The connection presented allows
  the design of electrokinetic experiments that correspond to particular active
  matter configurations that may, however, be  easier to conduct and
  control in the laboratory.
} \\

\end{tabular}

 \end{@twocolumnfalse} \vspace{0.6cm}

]
%%%END OF TITLE, AUTHORS AND ABSTRACT%%%

%%%FONT SETUP - please do not change any commands within this section
\renewcommand*\rmdefault{bch}\normalfont\upshape
\rmfamily
\section*{}
\vspace{-1cm}

%%%FOOTNOTES%%%

\footnotetext{\textit{1~School of Physics and Astronomy, University of
Minnesota, 116 Church St. SE, Minneapolis, MN 55455, USA.
E-mail:conk0044@umn.edu}}
\footnotetext{\textit{2~School of Physics and Astronomy, University of
Minnesota, 116 Church St. SE, Minneapolis, MN 55455, USA.
E-mail:vinals@umn.edu}}
\footnotetext{\textit{3~School of Physics and Astronomy, University of
Minnesota, 116 Church St. SE, Minneapolis, MN 55455, USA.
E-mail:otvalls@umn.edu}}

%Please use \dag to cite the ESI in the main text of the article.
%If you article does not have ESI please remove the the \dag symbol from the title and the footnotetext below.
%\footnotetext{\dag~Electronic Supplementary Information (ESI) available: [details of any supplementary information available should be included here]. See DOI: 10.1039/b000000x/}
%additional addresses can be cited as above using the lower-case letters, c, d, e... If all authors are from the same address, no letter is required

%\footnotetext{\ddag~Additional footnotes to the title and authors can be included \textit{e.g.}\ `Present address:' or `These authors contributed equally to this work' as above using the symbols: \ddag, \textsection, and \P. Please place the appropriate symbol next to the author's name and include a \texttt{\textbackslash footnotetext} entry in the the correct place in the list.}
%%%END OF FOOTNOTES%%%

%%%MAIN TEXT%%%%
\onecolumn

\section{Introduction}

Research on electrokinetic phenomena in liquid crystal nematics is currently
addressing the use of electrostatic fields to induce fluid flow, and to control
the motion of suspended particles. The anisotropic physical properties of the
liquid crystal molecules together with long range orientational order in a
nematic phase enable complex streaming flows in the bulk that can be controlled
by manipulating either the nematic director or applied electric fields.
Similarly, suspending self propelling particles (\lq\lq active matter'') in a
nematic matrix has been shown to allow control and steering of their motion by
designing appropriate nematic director configurations. In this paper, we advance
a formal analogy between these two distinct physical systems that allows the
extension, within certain limits, of results on transport in electrokinetic
systems to those for self propelled objects. Such a connection may allow the
design of electrokinetic experiments that are analogs of particular active
matter configurations of interest, and hence easier to conduct and control in
the laboratory.

The term electrokinetic phenomena refers collectively to induced response in
fluid electrolytes under imposed electrostatic fields, and to any resulting
fluid flow or suspended particle motion. Microscale manipulation of colloidal
particles and fluids by electric fields is a broad area of active scientific
research ranging from fundamental studies of non-equilibrium phenomena
\cite{re:bazant09,re:aranson13,re:zottl16,re:lavrentovich16} to the development
of practical devices for informational displays, portable diagnostics, sensing,
delivery, and cell sorting \cite{re:dobnikar13,re:ramos11,re:bazant10}.
Electrokinetic fluid transport is important in a variety of engineering, soft
matter, and biological systems. For example, electrokinetic flows have been used
to create \lq\lq lab on a chip'' micropumps, nanofluidic diodes, microfluidic
field effect transistors, and e-ink devices such as book
readers\cite{re:studer04,re:zhao12,re:comiskey98,re:klein13}.  Our specific
focus is on electrokinetic phenomena in the particular case in which the fluid
is a liquid crystal in the nematic phase.  Although ionic impurities are always
present in liquid crystal media, their effect has been usually considered as
parasitic, and thus to be minimized in applications. However, the recent
discovery of electrokinetic phenomena in nematic suspensions \cite{re:lazo14}
has opened a variety of avenues for the creation and control of designer flows
that rely on the anisotropy of the medium \cite{re:hernandez-navarro14}.

We explore here the mapping between the electrokinetic
problem just outlined, and that of the motion of self propelled
particles in a nematic matrix. In the latter case, the suspended
particles are endowed with an assumed speed (of internal
origin) along a preferred direction. When such particles are immersed
in a nematic matrix (a \lq\lq living liquid crystal''), they affect, and are
affected by, the orientational order in the matrix. Particles
move preferentially along the nematic direction in the matrix, both
because of their intrinsic velocity, and because of forces of
elastic origin imparted by the nematic medium. This motion of active particles
drives flow in living liquid crystals with a body force $\bm
f\sim\nabla\cdot(c\bm{nn})$, where $c$ is
the concentration of active particles and $\bm{n}$ is the nematic director \cite{re:simha02}.
We show that under certain conditions it is possible to map this physical
problem to the electrokinetic problem discussed above, and hence to use existing
results concerning flow induced by nematic director patterns to the case of
living liquid crystals. We develop this correspondence below, and study specific
configurations of interest in which the motion of active particles can be
controlled by imposed nematic director distributions.  The active
matter experiments of interest consist of a thin film of bacteria
(such as Bacillus subtilis) suspended in a lyotropic chromonic liquid
crystal matrix \cite{re:peng16a, re:genkin17}. Our modeling is
  confined to two dimensions. The experiments that we focus on, both in living
  liquid crystals and electrokinetic flows \cite{re:peng15}, 
  have been conducted
  in thin cells with patterned, fixed, director orientations. There is no
  indication in the experiments of any flow structure along the thin dimension,
  although the existence of no slip conditions on both top and bottom cell
  surfaces may introduce significant flow damping. For flows driven by
  director configurations involving isolated defects, damping may be important
  relative to viscous Newtonian stresses only at distances from the cores much
  larger than the cell thickness. In the case of Living Liquid Crystal
  experiments, the cell thickness is on the order of $5\mu$m, yet the swirling
  bacteria ensemble has its largest velocity around 35 $\mu$m. 
  On the other hand,
  cell ticknesses in electrokinetic experiments are much larger, on the order 
  of 50-100 $\mu$m \cite{re:peng15}. In order to highlight functional 
  dependencies of flow velocities with distance away from defects, and to
  compare our results with existing experiments and models, we confine our 
  analysis below to two dimensions, and neglect viscous damping terms. 
  However, it is straightforward to include them in the analysis as they
  are linear in the velocity. Finally, note that although we consider a
  two-dimensional problem, all densities introduced are defined per unit
  volume.

\section{Liquid crystal electrokinetics in an oscillatory electric field}

\subsection{Governing equations for the ionic system}

We consider a thin film of a liquid crystalline fluid, electrically
neutral, in its
nematic phase.  The fluid contains two ionic species of charge $\pm
e$, where $e$ is the elementary (positive) charge. It is
subjected to an external electrostatic field, 
spatially uniform but oscillatory in time, $\bm{E}_{0}$. The equations governing the evolution of the system
include species mass conservation, momentum
conservation in the fluid, electrostatic equilibrium, and torque
balance on the liquid crystal molecules \cite{re:conklin17}. Species mass conservation reads,
\begin{equation}
\frac{\partial c_{k}}{\partial t} + \nabla \cdot(\bm{v} c_{k}) =
\nabla \cdot \left( \bm{D} \cdot \nabla c_{k} - c_{k} z_k \bm{\mu}
  \cdot \bm{E} \right),
\label{eq:concentration}
\end{equation}
where $c_{k}, k=1,2$ is the concentration of species $k$, $z_1=1,z_2=-1$, $\bm{v}$ is
the barycentric velocity, which is equal to that of the liquid crystal
as the masses of the ions are negligible. The quantities $\bm{D}$ and
$\bm{\mu}$ are the mass diffusivity and ionic mobility tensors respectively, which will be
assumed to be anisotropic and depend on the local orientation of the
liquid crystalline molecule. They are also assumed to obey  Einstein's
relation $\bm{D} = ({k_{B}T}/{e}) \bm{\mu}$. 
The mobility tensor $\bm{\mu}$ is also assumed to be anisotropic, and to
depend on the local orientation of the nematic \cite{re:helfrich69} via
$\mu_{ij} = \mu_\perp\delta_{ij}+\Delta \mu \; n_i n_j$, where
$\delta_{ij}$ is the Kroenecker delta, and we define $\Delta\mu =
\mu_{\parallel}-\mu_{\perp}$, where $\mu_\parallel$ and $\mu_\perp$
are the ionic mobilities parallel and perpendicular to $\bm{n}$,
respectively. There is a great variety of possible electrokinetic
effects in a nematic suspension depending on physical parameters and
frequency of the applied fields. We focus on parameter ranges
suitable for experiments in electroosmotic flow and electrophoretic
motion as given, for example, in Peng, et al. \cite{re:peng15}. In particular,
we will focus on the limit of small anisotropy $\Delta \mu/\mu_\perp \ll 1$ ($\Delta\mu/\mu_{\perp} \approx 0.4$ in typical experiments
\cite{re:peng15}).

In the low frequency range of interest in electrokinetic experiments, the system is assumed to be in electrostatic equilibrium, so that the total electrostatic field in the medium satisfies
\begin{equation}
\label{eq:Poisson}
\epsilon_{0}\nabla \cdot ( \bm{\epsilon} \cdot \bm{E} )= \rho
\end{equation}
with charge density $\rho=e(c_1-c_2)$.
Although the liquid crystal molecules are not charged, they are polarizable \cite{re:degennes93}. The nematic is assumed to be a linear dielectric medium, with dielectric tensor $\epsilon_{ij} = \epsilon_{\perp}\delta_{ij} + \Delta\epsilon \; n_i n_j$, with $\Delta\epsilon = \epsilon_\parallel-\epsilon_\perp$, where $\epsilon_\parallel$ and $\epsilon_\perp$ are the dielectric constants parallel and perpendicular to $\bm{n}$, respectively. 

The liquid crystal is incompressible, $\nabla \cdot \bm{v} = 0$, and
flow is overdamped (typical Reynolds number $Re \sim 10^{-5} - 10^{-4}$). Momentum balance then reduces to the balance between the incompressible viscous stresses and the body forces exerted by the ionic species {and the nematic polarization} in a field\cite{re:degroot84,re:tovkach16},
\begin{equation}
\nabla \cdot \bm{T} + \rho\bm{E}+(\bm{D'}\cdot\nabla)\bm{E} = 0,
\label{eq:momentum_conservation}
\end{equation}
where $D'_{i}=\epsilon_0\epsilon_{ij}E_j$ is the electric displacement field. The stress tensor is $T_{ij} = -p \delta_{ij} + T_{ij}^{e} + \tilde T_{ij}$, where $p$ is the hydrostatic pressure and $\bm{T^e}$ is the elastic stress,
\begin{equation}
\label{eq:elastic_stress}
T_{ij}^{e} = -\frac{\partial f}{\partial (\partial_j
  n_k)}\frac{\partial n_k}{\partial x_i}
\end{equation}
with $f$ denoting the Oseen-Frank elastic free energy density
\cite{re:degennes93}. The viscous stress, $\bm{\tilde{T}}$, is
assumed to be given by the Leslie-Ericksen model \cite{re:degennes93}. 
The last term on the left hand side of Eq. ({\ref{eq:momentum_conservation}}) can be written as,
\begin{equation}
(\bm{D'}\cdot\nabla)\bm{E} = \nabla\left(\frac{1}{2}\epsilon_0\epsilon_\perp|\bm{E}|^2\right)+\epsilon_0\Delta\epsilon(\bm{n}\cdot\bm{E})(\bm{n}\cdot\nabla)\bm{E}.
\label{eq:polarization_force}
\end{equation}
The first term in Eq. ({\ref{eq:polarization_force}) contributes only to a change in pressure and does not affect the flow velocity. Thus with a redefinition of the pressure, Eq. ({\ref{eq:momentum_conservation}}) can be rewritten as,}
\begin{equation}
\nabla \cdot \bm{T} + \rho\bm{E}+\epsilon_0\Delta\epsilon(\bm{n}\cdot\bm{E})(\bm{n}\cdot\nabla)\bm{E}=0.
\label{eq:momentum2}
\end{equation} 
Eq. ({\ref{eq:Poisson}) implies the charge density is linear in the electric field{\cite{re:conklin17}}; thus both driving terms in Eq. ({\ref{eq:momentum2}}) are quadratic in the electric field {\cite{re:tovkach16}}, leading to persistent flow even in an AC field.}

Angular momentum conservation defines the dynamics of the director. A
torque balance argument yields \cite{re:degennes93}
\begin{equation}
\label{eq:directorDynamics}
\bm{n} \times \bm{h^{0}} - \bm{n} \times \bm{h}^{\prime} + \epsilon_0
\Delta \epsilon(\bm{n} \cdot\bm{E})(\bm{n} \times \bm{E})=0,
\end{equation}
where
\begin{equation}
\label{eq:h0}
h^{0}_i = -\frac{\partial f}{\partial n_i} + \frac{\partial}{\partial
  x_j} \frac{\partial f}{\partial (\partial_{j}
  n_i)}, \quad\quad
h^{\prime}_i = \gamma_1 N_i + \gamma_2 A_{ij}n_j,
\end{equation}
with $\gamma_1$ and $\gamma_2$ being rotational viscosities, $N_i = \dot n_i - W_{ij}n_j$, and $A_{ij} = \frac12 \left(
  \frac{\partial v_j}{\partial x_i} + \frac{\partial v_i}{\partial
    x_j} \right)$ and {$W_{ij} = \frac12 \left( \frac {\partial
    v_i}{\partial x_j}- \frac{\partial v_j}{\partial x_i} \right)$} the
symmetric and antisymmetric parts of the velocity gradient tensor. The
first term in Eq. (\ref{eq:directorDynamics}) corresponds to the
elastic torque on the director field, the second term corresponds to
viscous torque, and the third term is the torque due to the anisotropy
of nematic polarization.

\subsection{Variable orientation electric field as the analog of active
  stress}

We consider an imposed electric 
field that contains two orthogonal components of different frequency and phase, $\bm{E}_{0} = E_x\bm{\hat{x}}
\cos(\omega_x t) + E_y \bm{\hat{y}} \cos (\omega_y t + \psi)$. This field reduces
to a rotating field of constant magnitude when $E_x=E_y$, $\omega_x=\omega_y$, and
$\psi=\pi/2$.
We introduce dimensionless variables as follows: We scale the electric field by 
$E_x$ and time by $\omega_x^{-1}$; thus in dimensionless units, the applied field is 
$\bm{E}_{0} = \bm{\hat{x}}\cos t+A\bm{\hat{y}}\cos(\beta t+\psi)$, where $A=E_y/E_x$ and $\beta=\omega_y/\omega_x$.
We scale spatial variables by system size $L$, and the total ionic concentration $C=c_1+c_2$ by its average $c_0$. The scale of the charge density is {\cite{re:conklin17}} $\epsilon_0\epsilon_\perp E_x/L$, while the scale of the flow velocity and pressure are \cite{re:conklin17} $\epsilon_0\epsilon_\perp E_x^2 L/\alpha_4$ and $\epsilon_0\epsilon_\perp E_x^2$. The resulting set of dimensionless equations are,
\begin{equation}
\label{eq:C_dimensionless}
\Omega\frac{\partial C}{\partial t} + U \frac{\partial(C
  v_i)}{\partial x_i} = \gamma \frac{\partial}{\partial
  x_i}\left[\frac{D_{ij}}{D_\perp} \frac{\partial C}{\partial x_j}
\right] - Y^2 \frac{\partial}{\partial x_i} \left[ \rho
  \frac{\mu_{ij}}{\mu_\perp} E_{j} \right]
\end{equation}
\begin{equation}
\label{eq:rho_dimensionless}
\Omega \frac{\partial \rho}{\partial t} + U \frac{\partial (\rho
  v_i)}{\partial x_i} = \gamma \frac{\partial}{\partial x_i} \left[
  \frac{D_{ij}}{D_\perp} \frac{\partial \rho}{\partial x_j} \right] -
\frac{\partial}{\partial x_i} \left[ C\frac{\mu_{ij}}{\mu_\perp} E_{j} \right]
\end{equation}
\begin{equation}
\label{eq:poisson_dimensionless}
\frac{\partial }{\partial x_i} \left[ \frac{\epsilon_{ij}}{
    \epsilon_\perp} E_{j} \right] = \rho,
\end{equation}
\begin{equation}
\label{eq:momentum_dimensionless}
\nabla \cdot \bm{T} + \rho \bm{E}+\frac{\Delta\epsilon}{\epsilon_\perp}(\bm{n}\cdot\bm{E})(\bm{n}\cdot\nabla)\bm{E} = 0, \quad\quad \bm{T} =
-p\mathbb{I}+\frac{1}{\text{Er}}\bm{T}^e+\bm{\tilde{T}},
\end{equation}
\begin{equation}
\label{eq:director_dimensionless}
\bm{n} \times \bm{h}^0 - \text{Er} \left( \bm{n} \times \bm{h'} -
  \frac{\Delta\epsilon}{\epsilon_\perp} (\bm{n} \cdot \bm{E})(\bm{n}
  \times \bm{E}) \right) = 0,
\end{equation}
where $\Omega = \omega_x \tau_\rho$ is the driving
frequency relative to the charging time $\tau_{\rho}=\epsilon_{0}\epsilon_\perp
/(ec_{0}\mu_\perp)$,  {$U = \tau_\rho\epsilon_0\epsilon_\perp E_x^2/\alpha_{4}$},
$\gamma = \tau_\rho D_\perp/L^{2}$, {where $D_\perp$ is the ionic diffusivity perpendicular to $\boldsymbol{n}$}, $Y = \epsilon_0\epsilon_\perp E_x/(L
e c_0)$, and {$\text{Er}=\epsilon_0\epsilon_\perp E_x^2 L^{2}/K$} is the
Ericksen number, the ratio of viscous to elastic torques, {with $K$ as 
the average %otvfixed 
{value of the elastic constants} in the Oseen-Frank elastic free energy.} Note that
$\gamma$ can be also be written as $\gamma =
\lambda_D^{2}/L^{2}$, where $\lambda_D =
\sqrt{\epsilon_0\epsilon_\perp k_B T/(e^2 c_0)}$ is the Debye length. 
Also, in the scaled variables {$N_i = (\Omega/U)\partial_t
n_i+v_j\partial_j
n_i-W_{ij}n_j$.} Eqs. (\ref{eq:C_dimensionless}) - (\ref{eq:director_dimensionless})
represent the full set of governing equations in dimensionless form. 

{Consistent with typical electrokinetic experiments, we assume
that fluid anisotropy is small, and we expand the governing equations in powers of $\Delta\mu/\mu_\perp$ and $\Delta\epsilon/\epsilon_\perp$. At zeroth order in 
%otvfixed $\Delta$, 
{these two quantitites}, the equations correspond to a purely isotropic medium, with $C^{(0)}=1$, $\rho^{(0)}=0$, $\bm{v}^{(0)}=0$, and $\bm{E}^{(0)}=\bm{E}_0$.
Using Eq. ({\ref{eq:poisson_dimensionless}}), Eqs. ({\ref{eq:C_dimensionless}}) and ({\ref{eq:rho_dimensionless}}) at first order can be written as,}
\begin{equation}
\Omega\frac{\partial C^{(1)}}{\partial t} = \gamma \nabla^2 C^{(1)}-Y^2 (\bm{E}_0\cdot\nabla)\rho^{(1)}
\end{equation}
\begin{equation}\label{eq:rho_1}
\Omega \frac{\partial \rho^{(1)}}{\partial t} = \gamma \nabla^2\rho^{(1)} -\rho^{(1)}+ \left( \frac{\Delta
    \epsilon}{\epsilon_{\perp}} - \frac{\Delta \mu}{\mu_{\perp}}
\right) \nabla\cdot(\bm{n} (\bm{n} \cdot \bm{E}_{0}))-(\bm E_0\cdot\nabla)C^{(1)}.
\end{equation}
{Similarly, 
we assume a system in which $Y^2/(4\gamma\sqrt{1+\Omega^2})\ll 1$, which can be shown implies $C^{(1)}$ and $\rho^{(1)}$ decouple {\cite{re:conklin17}}. Furthermore, since the Debye length in electrokinetic systems is typically on the order of $\lambda_D\sim 10^{-6}$ m, while cell sizes are  $L\sim 10^{-4}$ to $10^{-3}$ m, we find $\gamma\sim 10^{-6}$ to $10^{-4}$. Thus the diffusion term in Eq. ({\ref{eq:rho_1}}) is negligible far from nematic defect cores{\cite{re:conklin17}}. Therefore Eq. {(\ref{eq:rho_dimensionless}}) can be written to first order in 
the anisotropies as,}
\begin{equation}
\label{eq:rho}
\Omega \frac{\partial \rho}{\partial t} + \rho = \left( \frac{\Delta
    \epsilon}{\epsilon_{\perp}} - \frac{\Delta \mu}{\mu_{\perp}}
\right) \nabla\cdot(\bm{n} (\bm{n} \cdot \bm{E}_{0})).
\end{equation}
The solution to Eq. (\ref{eq:rho}) is given by,
\begin{equation}
\rho(\bm{r},t) = \left( \frac{\Delta
    \epsilon}{\epsilon_{\perp}} - \frac{\Delta \mu}{\mu_{\perp}}
\right) \frac{\cos(t-\delta)}{\sqrt{1+\Omega^2}} \nabla \cdot( \bm{n}
n_x) + \left( \frac{\Delta
    \epsilon}{\epsilon_{\perp}} - \frac{\Delta \mu}{\mu_{\perp}}
\right)  \frac{ \cos(\beta t +
  \psi-\delta_2)}{\sqrt{1+(\beta\Omega)^2}} A \nabla\cdot( \bm{n} n_y),
\end{equation}
with $\tan \delta = \Omega$ and $\tan\delta_2 = \beta\Omega$. 

{Since the applied field $\bm{E}_0$ is spatially uniform, the body force on the fluid due to nematic polarization (the last term on the left hand side of Eq. ({\ref{eq:momentum_dimensionless}})) is second order in the anisotropies. To first order in $\Delta$, the body force on the nematic fluid is therefore $\bm{f} \approx \rho
\bm{E}_0 =\rho (\bm{r}, t) (\bm{\hat{x}} \cos t + A \bm{\hat{y}} \cos(
\beta t + \psi))$, or}
\begin{eqnarray}
\bm{f} & = &   \left( \frac{\Delta
    \epsilon}{\epsilon_{\perp}} - \frac{\Delta \mu}{\mu_{\perp}}
\right) \left[ \frac{\cos t \cos(t-\delta)}{\sqrt{1+\Omega^2}} \nabla
  \cdot(\bm{n} n_x \bm{\hat{x}} ) + \frac{A^2\cos(\beta t + \psi)
    \cos (\beta t + \psi - \delta_2)}{\sqrt{1+(\beta\Omega)^2}}
  \nabla \cdot(\bm{n} n_y \bm{\hat{y}} )\right. \nonumber \\
& & \left. + \frac{A \cos (\beta t + \psi) \cos(t -
    \delta)}{\sqrt{1+\Omega^2}} \nabla \cdot( \bm{n} n_x
\bm{\hat{y}}) + \frac{A \cos t \cos(\beta t + \psi -
  \delta_2)}{\sqrt{1+(\beta\Omega)^2}} \nabla \cdot(\bm{n} n_y
\bm{\hat{x}}) \right]
\label{eq:f}
\end{eqnarray}

We now define a time-averaged force, {$\langle\bm f\rangle = \lim\limits_{T\rightarrow\infty}(1/T)\int_0^T\bm f dt$}. The time averages are performed using:
\begin{equation}
\lim_{T\rightarrow\infty}\frac{1}{T}\int_0^T\cos(t-\delta)\cos(\beta t+\psi)dt = \left\{\begin{array}{cc}\frac{1}{2}\cos(\delta+\psi),& |\beta| = 1 \\0,& \text{otherwise}\end{array}\right.
\end{equation}
\begin{equation}
\lim_{T\rightarrow\infty}\frac{1}{T}\int_0^T\cos t\cos(\beta t+\psi-\delta_2)dt = \left\{\begin{array}{cc}\frac{1}{2}\cos(\delta-\psi),& |\beta| = 1 \\0,& \text{otherwise}\end{array}\right.
\end{equation}
Assume first that $|\beta| \neq 1$. Then the last two terms of Eq. (\ref{eq:f}) average to zero, and the average force is,
\begin{equation}
\langle \bm{f} \rangle = \left( \frac{\Delta
    \epsilon}{\epsilon_{\perp}} - \frac{\Delta \mu}{\mu_{\perp}}
\right) \left[ \frac{\nabla\cdot( \bm{n} n_x \bm{\hat{x}})
  }{2(1+\Omega^2)} + \frac{A^2 \nabla \cdot(\bm{n} n_y \bm{\hat{y}})
  }{2(1+(\beta\Omega)^2)} \right]
\label{eq:avg_2}
\end{equation}
For the specific choice of $A=
\sqrt{(1+(\beta\Omega)^2)/(1+\Omega^2)}$, so that with
\begin{equation}
\label{eq:E_choice}
\bm{E}_{0} =\bm{\hat{x}} \cos t + \bm{\hat{y}} \sqrt
{\frac{1+(\beta\Omega)^2}{1+\Omega^2}}\cos(\beta t + \psi),
\end{equation} 
for arbitrary $\psi$ and $\beta\neq 1$, Eq. (\ref{eq:avg_2}) becomes,
\begin{equation}
\langle \bm{f} \rangle =  \left( \frac{\Delta
    \epsilon}{\epsilon_{\perp}} - \frac{\Delta \mu}{\mu_{\perp}}
\right) \frac{\nabla\cdot(\bm{n} \bm{n})}{2(1+\Omega^2)}.
\label{eq:avg_active}
\end{equation}
Equation (\ref{eq:avg_active}) has the same form as the driving force in active
nematics when the concentration of swimmers is constant \cite{re:simha02,re:green16}. 
Thus our analysis predicts active-like flows on average in electrokinetic systems
driven by the field of Eq. (\ref{eq:E_choice}). If $\beta =1$,
the force will not be active-like unless the director is fixed in specific configurations \cite{re:conklin17PhD}.

\section{Results}

To illustrate features of active-like motion that would take place in
the liquid crystal electrokinetic analog, we 
numerically investigate electrokinetic flows generated by the electric field 
of Eq. (\ref{eq:E_choice}).
The experimental configuration that we have in mind involves a thin
film (tens of microns in thickness) of a nematic liquid
crystal with tangential anchoring on top and bottom surfaces (director
parallel to the surface). A photosensitive material is coated onto
the plates bounding the film, which are then exposed to light that has
been polarized through a mask with nanoslits etched in the desired
director pattern \cite{re:peng15,re:guo16}. This exposure aligns the
primary axes of photo sensitive molecules with the desired pattern
\cite{re:yaroshchuk12}. For sufficiently thin films, the
photopatterned director field is largely constant, uniform across the film.
Lithographic surface patterning offers
the opportunity of tailoring flow fields in nematics for specific
applications, for example, to engineer flows in microfluidic channels,
or to effect controlled immersed particle motion or species
separation. It is also possible to design on demand director patterns
which can be reconfigured dynamically during an experiment
\cite{re:guo16}. Such a configuration has been used recently to
control the motion of bacteria in a lyotropic liquid crystal
\cite{re:peng16b}. 

The governing equations are integrated numerically with the commercial
software package COMSOL. The solutions were obtained on a circular
domain $C_0$ with radius $r_0=1$. Within $C_0$ is a second circular
domain $C_1$ with radius $r_1=1/5$, in which the mesh is more finely
resolved. $C_1$ contains 109,196 triangular elements of linear size
between $6.4\times10^{-6}$ and $1.8\times 10^{-3}$, while $C_0$
contains 12,790 elements with linear size between $1.6\times10^{-4}$
and $0.13$. Additionally, $C_0$ contains 384 quadrilateral elements to
resolve the boundary layer at $r=r_1$. Equations
(\ref{eq:C_dimensionless}) through (\ref{eq:momentum_dimensionless})
are iterated in time, while the director field $\bm{n}(\bm{r})$ is
held fixed. The solution is obtained with no flux boundary conditions
for the concentrations, no slip boundary conditions for the velocity, and
Dirichlet boundary conditions $\Phi = -x\cos(t)-yA\cos(2 t)$ for the electric potential, with $A$ satisfying Eq. (\ref{eq:E_choice}) above. {The numerical solutions assume $\Delta\epsilon/\epsilon_\perp=0$ and $\Delta\mu/\mu_\perp=0.4$, consistent with recent electrokinetic experiments{\cite{re:peng15}}}. Further details of the numerical method have been discussed elsewhere \cite{re:conklin17, re:conklin17PhD}.
We present next the results for two director configurations which have been studied in living liquid 
crystal experiments: a single fixed point defect, and a pair of point
defects with opposite topological charge \cite{re:peng16b}. 

\begin{figure}
\centering
\includegraphics[width=0.4\textwidth]{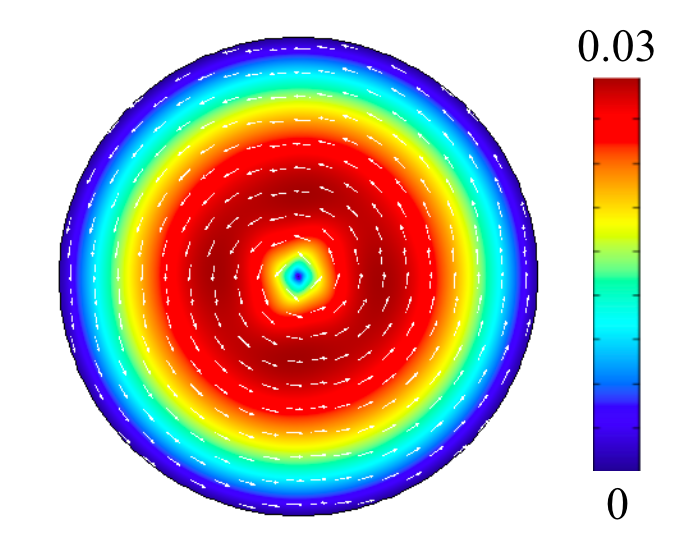}
\includegraphics[width=0.5\textwidth]{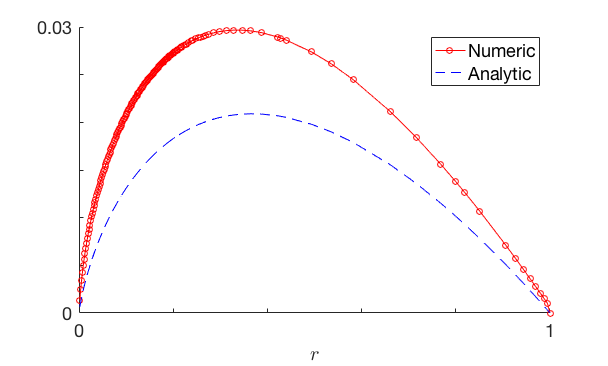}
\caption{Left: Average electrokinetic velocity corresponding to an imposed director field $\theta=\phi-\pi/4$, and a two frequency
  applied field with %otvfixed applied field with 
$\beta = 2, \psi =
  0$, {obtained by numerically integrating Eqs. ({\ref{eq:C_dimensionless}}) through ({\ref{eq:momentum_dimensionless}})}. Right: radial velocity profile from the figure compared with the
  analytic solution, Eq. (\ref{eq:spiral_analytic_v}).}
\label{fig:SpiralTwoFreq}
\end{figure}

Figure \ref{fig:SpiralTwoFreq} shows the numerically computed average
velocity for the electrokinetic model when the director pattern is
given by single (+1) defect of director field $\bm{n}(\bm{r}) = (\cos \theta(\bm{r}),
\sin \theta(\bm{r}))$ with $\theta(r,\phi) = \phi - \pi/4$ at the
center of the computational domain. The angle $\phi$ is the azimuth in
polar coordinates. The constant phase $-\pi/4$
creates the {vortex with swirling arms} studied in experiments of living liquid crystals\cite{re:peng16b}. Interestingly,
the velocity field is not parallel to the local nematic, as noted in
the experiments. Whereas
this is surprising in the context of a living liquid crystal in which
bacteria are known to move parallel to the local director, it is not so
for an electrokinetic system. In the latter case, motion is due to the
local body force that originates from charge separation, and does not
in general follow director lines. Instead, charge accumulates in
regions in which the director is normal to the imposed electric field.

The velocity field obtained can be computed analytically by assuming a Newtonian fluid. Consider a
director field comprising a single (+1) defect $\theta(r,\phi) = \phi + \alpha$, 
where $\alpha$ is an arbitrary constant phase. Equation (\ref{eq:avg_active}) becomes,
\begin{equation}
\label{eq:disc_force}
\langle \bm{f} \rangle = \left( \frac{\Delta
    \epsilon}{\epsilon_{\perp}} - \frac{\Delta \mu}{\mu_{\perp}}
\right)\frac{\cos(2\alpha)}{2r(1+\Omega^2)}\bm{\hat r}+\left( \frac{\Delta
    \epsilon}{\epsilon_{\perp}} - \frac{\Delta \mu}{\mu_{\perp}}
\right)\frac{\sin(2\alpha)}{2r(1+\Omega^2)}\bm{\hat{\phi}}.
\end{equation}
The first term in Eq. (\ref{eq:disc_force}) may be written as $\nabla g$, where
$$ g = \left( \frac{\Delta
    \epsilon}{\epsilon_{\perp}} - \frac{\Delta \mu}{\mu_{\perp}}
\right)\frac{\log r\cos(2\alpha)}{2(1+\Omega^2)},
$$
and therefore this term can be included in the pressure field of the
incompressible fluid. The second term in Eq. (\ref{eq:disc_force}) also has a nonzero curl and can be rewritten as,
$$
\left( \frac{\Delta
    \epsilon}{\epsilon_{\perp}} - \frac{\Delta \mu}{\mu_{\perp}}
\right)\frac{\sin(2\alpha)}{2r(1+\Omega^2)}\bm{\hat{\phi}}=\nabla\left[\left( \frac{\Delta
    \epsilon}{\epsilon_{\perp}} - \frac{\Delta \mu}{\mu_{\perp}}
\right)\frac{\phi\sin(2\alpha)}{2(1+\Omega^2)}\right].
$$
However, if this term were included in the pressure, we would find that the
pressure is not single valued, $p(\phi)\neq p(\phi+2\pi)$, which is unphysical.
Therefore the body force given by the second term Eq. (\ref{eq:disc_force}),
though irrotational, cannot be included in the pressure, and must ultimately be
balanced by a viscous force instead. If we assume the viscous stress to be
Newtonian, $\tilde{T}_{ij}=\partial_j v_i$, an averaged momentum balance in two
spatial dimensions can be written as,

\begin{equation}
\label{eq:NS}
-\nabla p' + \nabla^2 \bm{v} -\zeta^2 \bm{v}+ \left( \frac{\Delta
    \epsilon}{\epsilon_{\perp}} - \frac{\Delta \mu}{\mu_{\perp}}
\right)\frac{\sin(2\alpha)}{2r(1+\Omega^2)}\bm{\hat{\phi}}= 0, \quad\nabla\cdot\bm v=0,
\end{equation}
where
$$
p' = p -  \left( \frac{\Delta
    \epsilon}{\epsilon_{\perp}} - \frac{\Delta \mu}{\mu_{\perp}}
\right)\frac{\log r\cos(2\alpha)}{2(1+\Omega^2)}.
$$
{The damping term $-\zeta^2 \bm{v}$, $\zeta = 2\sqrt{3}L/h$,  arises from depth-averaging the velocity profile, assuming a Poiseuille flow in the $z$ direction} \cite{re:genkin17,re:thom89}. {Here $h$ is the cell thickness.}

The solution to Eq. (\ref{eq:NS}) in a disc of dimensionless radius 1, 
with no-slip boundary conditions is constant $p'$ and
\begin{equation}
\label{eq:v_bessel}
\bm{v}=\left(\frac{\Delta\epsilon}{\epsilon_\perp}-\frac{\Delta\mu}{\mu_\perp}\right)\frac{\bm{\hat{\phi}} \sin(2\alpha)}{2 \zeta(1+\Omega^2)}\left[ \frac{1}{r\zeta}-K_1(r\zeta)+\frac{[\zeta  K_1(\zeta )-1]
   }{\zeta  I_1(\zeta )}I_1(r \zeta )\right],
\end{equation}
{where $I_1,K_1$ are modified Bessel functions of the first and second kind,
respectively. When $r\zeta\gg 1$ the flow is exponentially damped. 
In the opposite limit,} 
\begin{equation}
\label{eq:v_bessel_limit}
\bm{v}(r\zeta\ll 1) \approx -\frac{1}{4}\left(\frac{\Delta\epsilon}{\epsilon_\perp}-\frac{\Delta\mu}{\mu_\perp}\right)\frac{\bm{\hat{\phi}} \sin(2\alpha)}{1+\Omega^2}[r\log r-r \eta(\zeta)],
\end{equation}
where
$$
\eta(\zeta) = \frac{\zeta K_1(\zeta)-1}{\zeta I_1(\zeta)}+\log\left(\frac{2}{\zeta}\right)+\frac{1}{2}-\gamma,
$$
{with $\gamma$ being Euler's constant. In the special case in which $\zeta\ll 1$, $\eta(\zeta)\rightarrow 0$ and Eq.} (\ref{eq:v_bessel_limit}) {reduces to,}
\begin{equation}
\label{eq:v_spiral}
\bm{v} = - \frac{1}{4}\left( \frac{\Delta
    \epsilon}{\epsilon_{\perp}} - \frac{\Delta \mu}{\mu_{\perp}}
\right) \frac{ \bm{\hat{\phi}} \sin(2\alpha)r\log r}{1+\Omega^2}.
\end{equation}
In the specific case of $\alpha = - \pi/4$, one finds,
\begin{equation}
\bm{v} =  \frac{1}{4} \left( \frac{\Delta
    \epsilon}{\epsilon_{\perp}} - \frac{\Delta \mu}{\mu_{\perp}}
\right) \frac{r \ln r}{1+\Omega^{2}} \;  \bm{\hat{\phi}},
\label{eq:spiral_analytic_v}
\end{equation}
Equation (\ref{eq:spiral_analytic_v}) is compared in Fig.
\ref{fig:SpiralTwoFreq} to a fully numerical solution of the governing
equations. {While there is a noticeable difference in magnitude between the
two solutions, which is due to the additional approximations involved in the 
analytic solution,} Eq. (\ref{eq:spiral_analytic_v}) (Newtonian stress, small
anisotropy, etc.), both clearly exhibit an $r\log r$ dependence 
along $\bm{\hat\phi}$. {This result is in agreement with the velocity profile
reported by Peng, et al}\cite{re:peng16b} {for a living liquid crystal under
the same fixed director configuration (Fig. 2D shows the experimentally
determined azimuthal velocity profile), even though the experiments are
nominally conducted in the limit $h/L \ll 1$. }

%{It is possible to
%add a damping force $- \zeta {\bm v}$ to the left hand side of}
%Eq. (\ref{eq:NS}), {and still obtain an analytic solution. It difers from Eq.
%}
%(\ref{eq:v_spiral}) {only far from the defect core. We do not give it here 
%as there is not enough information in Ref.} 
%\cite{re:peng16b} {(Fig. 2d) to make any meaningful analysis.}

We examine next a dipolar configuration comprising a (+1/2) and a
(-1/2) defect pair. In the electrokinetic analog, the dipolar
nature of the configuration is expected to lead to nonzero average
flow, and directed from the (-1/2) defect
towards the (+1/2) defect.
\begin{figure}
\centering
\raisebox{3ex}{
\includegraphics[width=0.29\textwidth]{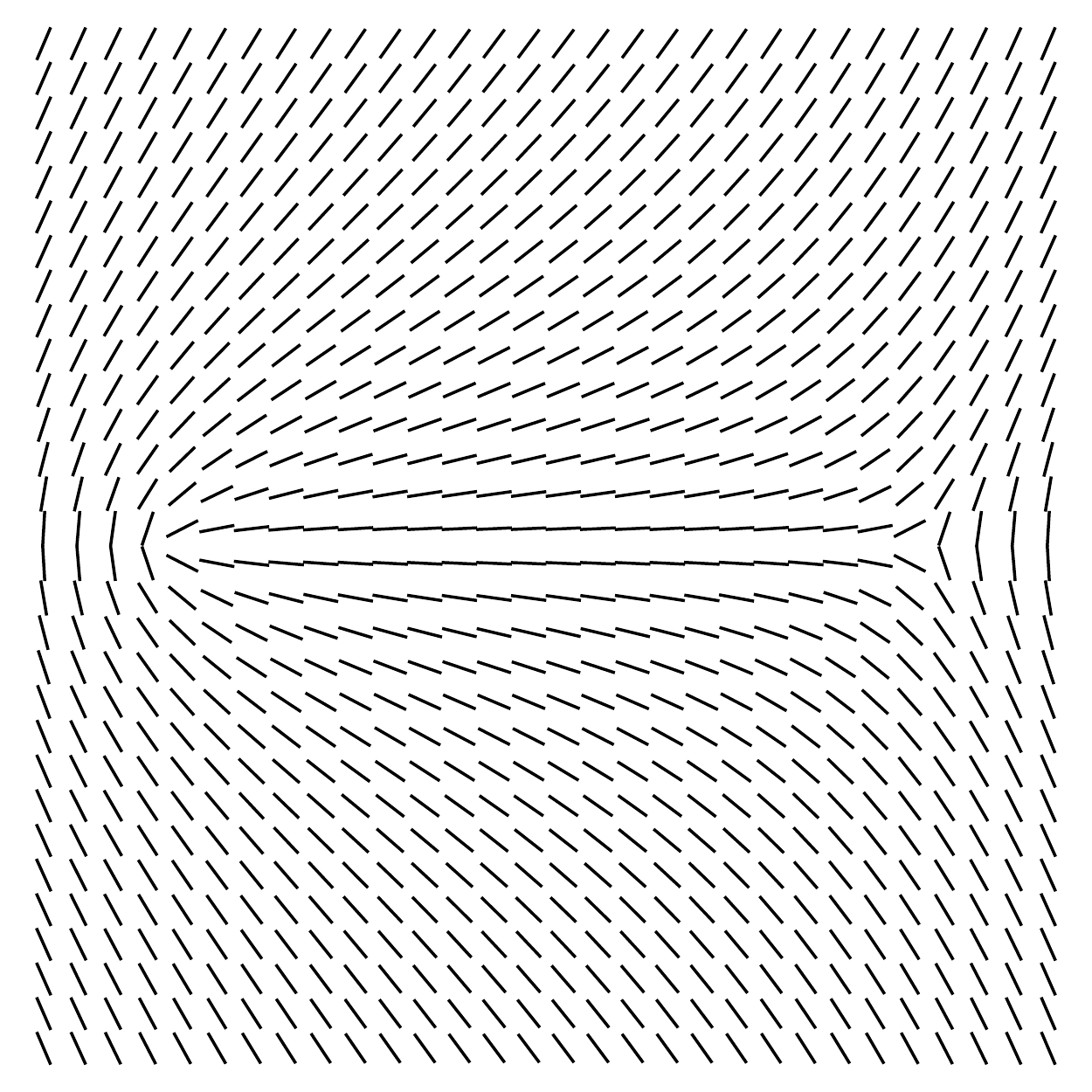}
}
\includegraphics[width=0.6\textwidth]{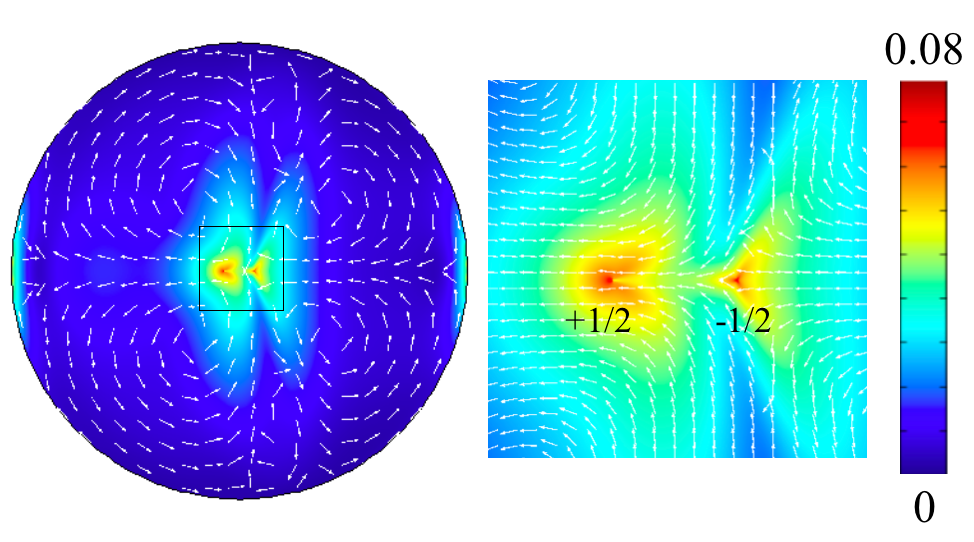}
\caption{Left: imposed director pattern comprising a (+1/2,-1/2)
  defect pair, as in the experiments of
  Ref. \cite{re:peng16b}. Center: Numerically determined fluid velocity,
  averaged over time. As is the case in the experiments, fluid flows
  from the $(-1/2)$ defect (high splay) towards the $(+1/2)$
  defect (high bend) {The velocity is very small near the top and bottom of the domain, leading to numerical noise in the plotting of the directional arrows}. Right: central region of the velocity map magnified.}
\label{fig:phmhTwoFreq}
\end{figure}
Figure \ref{fig:phmhTwoFreq} shows our results for the average
numerical electrokinetic velocity. This flow is also in agreement
with is the experiments in living liquid crystals
\cite{re:peng16b}. In that case, it is interpreted as flow originating
in regions of mixed splay-bend distortion from regions of high splay
to high bend.

\section{Discussion and conclusions}

Despite different underlying physical mechanisms, we find that forces
and corresponding flows in liquid crystal electrokinetic systems
behave on average like those of living liquid crystals when the
applied field has the form of Eq. (\ref{eq:E_choice}). The driving
force on an element of volume in the active system is due to the
self-propulsion of bacteria along $\pm\bm{n}$ and the corresponding
motion of the nematic in the opposite direction due to Newton's Third
Law \cite{re:simha02}. On the other hand, an element of volume in
liquid crystal electrokinetics is driven by the electrostatic force on
that element of volume. The amount and sign of charge in an element of
volume is given by the anisotropy of the nematic medium. By applying a
two component electric field with unequal frequencies and averaging
over time, the net electrostatic force is along $\pm\bm{n}$ as
well. This prediction has not yet been experimentally verified, and it
remains to be seen whether higher order corrections in anisotropy and
mass diffusion neglected in the analysis are indeed negligible in experimental situations.

The correspondence as derived above assumes uniform ionic and bacterial concentrations in electrokinetic and active systems, respectively. While some active nematic systems exhibit near-uniform concentration \cite{re:green16},   there are other cases in
which variations in  concentration are not negligible.
In particular, the experiments of interest \cite{re:peng16a} show significant variations in bacterial
concentration; the bacteria form an annulus in the {swirling vortex}
configuration, while in the defect dipole the bacteria cluster at the
(+1/2) defect and avoid the (-1/2) defect. In the electrokinetic system, one must account for the nonlinear coupling between $\rho$ and $C$ in Eqs. (\ref{eq:C_dimensionless}) and (\ref{eq:rho_dimensionless}) in order to determine whether the average body force $\langle\rho\bm{E}\rangle$ is active-like when variations in $C$ are not negligible. 
 
Additionally, charge separation induces an electric field of first order 
in the anisotropies, %$\Delta$,
 which may lead to unique flows when anisotropy is not small.  Furthermore,  the bacteria suspended in the nematic are typically several microns long, and thus cannot be assumed to be point particles as the ions in electrokinetic systems are. Thus, unlike ions, the bacteria in living liquid crystals are expected to distort the nematic orientation -- an effect which is not captured by the electrokinetic analogy.

Finally we note a few similar features between the evolution of ionic and bacterial concentrations in the two systems. We follow the analysis of Genkin, et al.\cite{re:genkin17} that
introduced two concentrations $c_{\pm}$ that denote separate
bacterial populations that swim with velocity $\bm{v}_0$ along the two possible directions
parallel to the local director $\bm{n}$. Each bacterial population
satisfies the equation of diffusion-advection, but can switch
orientation over a reversal time scale $\tau$,
\begin{equation}
\frac{\partial c_{\pm}}{\partial t} + \nabla \cdot (\pm v_{0} \bm{n}
c_{\pm} + \bm{v} c_{\pm} ) = D \nabla^{2} c_{\pm} -
\frac{c_{\pm}-c_{\mp}}{\tau}.
\label{eq:conservation_active}
\end{equation}
On the other hand, using Poisson's equation and assuming $\Delta\epsilon=0$, the conservation of ions in an electrokinetic system, Eq. (\ref{eq:concentration}), may be written as,
\begin{equation}
\frac{\partial c_k}{\partial t} + \nabla\cdot(z_k \Delta\mu(\bm{n}\cdot\bm{E})\bm{n} c_k+\bm{v} c_k)
= \nabla\cdot(\bm{D}\cdot\nabla c_k) - \left(\frac{\mu_\perp e c_k}{\epsilon_\perp\epsilon_0}\right)z_k(c_1-c_2)-z_k\mu_\perp(\bm{E}\cdot\nabla)c_k
\label{eq:conservation_ion}
\end{equation}

The most significant physical difference between the bacterial concentrations $c_\pm$ and ionic concentrations $c_k$ is that the ionic species $c_k$ are physically distinct and must be conserved, while only the total bacterial concentration $c_++c_-$ must be conserved. Nevertheless, we find a number of similarities between Eqs. (\ref{eq:conservation_active}) and (\ref{eq:conservation_ion}).  The anisotropy of ionic mobility leads to ionic drift along $\bm{n}$ with velocity $\Delta\mu(\bm{n}\cdot\bm{E})$, similar to bacterial self-propulsion. The ionic charging time $\epsilon_\perp \epsilon_0/(\mu_\perp e c_k)$ is analogous to the bacterial reversal time $\tau$, though the charging time is a function of local concentration $c_k$. The last term on the right hand side of Eq. (\ref{eq:conservation_ion}) is the only term without an analogous term in Eq. (\ref{eq:conservation_active}). Thus we find that the equations of bacterial concentration have a similar form to the equations of ionic concentration, though the ionic flux terms contain higher order nonlinearities than their bacterial analogs.

To summarize, we find a connection between flows in living liquid crystals and electrokinetic flows in nematics driven by a two-component oscillating field. While having different physical mechanisms, both systems drive flow with a force $\bm{f}\sim\nabla\cdot(\bm{nn})$. We compare experimental living liquid crystal flows with a fixed director pattern and numerical studies of the corresponding electrokinetic systems. While the numerical electrokinetic results show agreement with the observed bacterial flows, an experimental verification of this correspondence has not yet been performed. This mapping may 
be useful in using singularity solutions already known for liquid crystal electrokinetics to interpret singularity driven flows in active systems. It may also prove useful in that the experiments involving ionic systems are free of some of the complication inherent in handling active matter, including controlling the activity during the experiments. In this respect, 
the study of electrokinetic flows may become a tool in studying synthetic configurations involving designer flows, later to be verified directly on the living liquid crystal. 

\section{Conflicts of interest}
There are no conflicts of interest to declare.

\section{Acknowledgments}

We are indebted to Oleg Lavrentovich for introducing us to the subject of Living
Liquid Crystals, and for sharing experimental data. We are also indebted to
Chandan Dasgupta and Peter Stoeckl for many stimulating discussions. This research has been 
supported by the National Science Foundation under contract DMS 1435372, by the
Minnesota Supercomputing Institute, and by the Extreme Science and Engineering
Discovery Environment (XSEDE) \cite{re:towns14}, which is supported by National
Science Foundation grant number ACI 1548562. 

%%%REFERENCES%%%
\bibliography{references}
\bibliographystyle{rsc} %the RSC's .bst file

\end{document}